\documentclass[fleqn,usenatbib]{mnras}

\usepackage{newtxtext,newtxmath}
\usepackage{makecell}
\usepackage{float}

\usepackage[T1]{fontenc}

\DeclareRobustCommand{\VAN}[3]{#2}
\let\VANthebibliography\thebibliography
\def\thebibliography{\DeclareRobustCommand{\VAN}[3]{##3}\VANthebibliography}

\usepackage{graphicx}	
\usepackage{amsmath}	

\definecolor{tangerine}{rgb}{0.1, 0.6, 0.3}
\definecolor{forestgreen}{RGB}{34,139,34}

\title[Robust completeness test for $H_{0}$ with GWs]{Implementing a Robust Test of Galaxy Catalogue Completeness for Dark Siren Measurements of the Hubble Constant}

\author[L. E. H. Datrier et al.]{
Laurence E. H. Datrier,$^{1,2}$\thanks{E-mail: laurence.datrier@glasgow.ac.uk}
Martin A. Hendry$^{1}$
\\
$^{1}$School of Physics \& Astronomy, University of Glasgow, Glasgow G12 8QQ, UK\\
$^{2}$The Nicholas and Lee Begovich Center for Gravitational-Wave Physics and Astronomy, California State University, Fullerton, 92831, USA\\}

\date{Accepted XXX. Received YYY; in original form ZZZ}

\pubyear{\the\year{}}

\begin{document}
\label{firstpage}
\pagerange{\pageref{firstpage}--\pageref{lastpage}}
\maketitle

\begin{abstract}

We present the application of a robust test of galaxy catalogue completeness to the \texttt{gwcosmo} pipeline. The method implements a straightforward statistical test for determining the apparent magnitude completeness limit of a magnitude-redshift sample. This offers an improved, less conservative approach compared with how galaxy catalogue completeness is currently estimated in the \texttt{gwcosmo} gravitational wave cosmology pipeline for determining the Hubble constant $H_{0}$. The test also does not require prior knowledge of the luminosity function, and thus returns a more robust estimate of the limiting apparent magnitude for a magnitude-redshift sample of galaxies. For GWTC-1 results using $B$-band photometry of galaxies in the GLADE catalogue, we find a  $1.3\%$ improvement on the inference of $H_{0}$ using dark sirens only and a $3.4\%$ improvement for the combined posterior with GW170817. Using GLADE+, there is a $8.6\%$ improvement with dark sirens only and a $6.3\%$ improvement for the combined posterior with GW170817. However, the final posterior on $H_{0}$ using the GWTC-3 dataset with the GLADE+ $K$-band shows no improvement when applying the robust method. This is because the GLADE+ galaxy catalogue provides little or no coverage in the $K$-band for any of the GWTC-3 events. However, with the use of deeper galaxy catalogues in future gravitational wave cosmology analyses, the adoption of a less conservative estimate of magnitude completeness will become increasingly important.
\end{abstract}

\begin{keywords}
gravitational waves -- cosmological parameters -- methods: data analysis -- catalogues
\end{keywords}



\section{Introduction}

The precise value of the Hubble constant, a measure of the cosmic expansion rate, is currently a major point of contention in modern cosmology. At the time of writing, the discrepancy between early- (CMBR) and local- Universe measurements of $H_{0}$ has gone past 5$\sigma$ \citep{Planck2018,Riess_2022}. The latest Planck measurements give a value of $H_{0} = 67.4 \pm 0.5 {\text{\,km\,s}}^{-1}{\text{\,Mpc}}^{-1}$ \citep{Planck2018}, while the latest SHoES analysis yields $H_{0} = 73.04 \pm 1.04 {\text{\,km\,s}}^{-1}{\text{\,Mpc}}^{-1}$ \citep{Riess_2022}.

Over the past decade, gravitational-wave (GW) cosmology has emerged as a potentially powerful tool for resolving the current tension between early- and local- Universe measurements of $H_{0}$~\citep{Chen_2018, Borhanian_2020, 10.1093/mnras/stad2115, Bertheas:2025mzd}.
One method for determining the Hubble constant from GW signals is the so-called "galaxy catalogue method", which was first proposed in~\cite{Schutz}. Compact binary coalescences (CBCs) are self-calibrating distance indicators, yielding absolute distance measurements from analysis of their waveforms; they are referred to as standard sirens, the GW analogues to standard candles~\citep{Holz_2005}. Redshift information is degenerate with chirp mass in CBCs; therefore, their redshift $z$ must be obtained through other means. There are a number of methods for obtaining redshift information associated with dark sirens, several of which have been applied to real or simulated data to investigate the efficacy of bright and dark sirens for constraining the Hubble constant with current or future cosmological data: dark sirens~\citep{PhysRevD.86.023502, PhysRevD.77.043512,PhysRevD.96.101303,Soares-Santos_2019,Fishbach_2019,Palmese_2020, Gray_2023}, bright sirens~\citep{Chen_2018}, spectral sirens~\citep{1993ApJ...411L...5C,Farr_2019, PhysRevLett.108.091101}, cross-correlations~\citep{PhysRevD.103.043520, Mukherjee:2022afz,Afroz:2024joi}.
Here, we focus on the use of electromagnetic galaxy catalogues to obtain redshift information for GW events. Where a host galaxy cannot be identified through the detection of an electromagnetic counterpart to the GW event, one can statistically infer $H_{0}$ through the use of a galaxy catalogue. By assigning a probability to the potential host galaxies for each event, marginalising over these potential host galaxies, and combining the results for many events, a precise value for $H_{0}$ can be obtained \citep{Schutz, Del_Pozzo_2012}.

The success of the LIGO--Virgo--KAGRA (hereafter LVK) network of ground-based GW interferometers in detecting signals from compact binary mergers has made GW cosmology using CBCs as standard sirens a reality. In particular, the detection of the binary neutron star GW170817 and its electromagnetic counterpart provided our best individual constraint so far on $H_{0}$ from standard sirens, with an initial result of $H_{0} = 70.0^{+12.0}_{-8.0} {\text{\,km\,s}}^{-1}{\text{\,Mpc}}^{-1}$ \citep{BNShubble}. To date, the best LVK constraints on $H_{0}$ from combining 
standard sirens up to the end of O4a is $H_{0} = 76.6^{+13.0}_{-9.5} {\text{\,km\,s}}^{-1}{\text{\,Mpc}}^{-1}$ 
\citep{theligoscientificcollaboration2025gwtc40constraintscosmicexpansion}. Constraints from GWTC-3 give $H_{0} = 68^{+8}_{-6} {\text{\,km\,s}}^{-1}{\text{\,Mpc}}^{-1}$ \citep{Abbott_2023}.

While bright sirens --- i.e. standard sirens with EM counterparts --- remain our most powerful tool for inferring the value of $H_{0}$ from GW observations, only one such event has been detected to date~\citep{LIGOScientific:2017ync, BNShubble}. We must therefore make use of dark sirens to refine the posterior on $H_{0}$~\citep{Abbott_2023}.

A full description of the galaxy catalogue method for inferring $H_{0}$ from dark sirens, as implemented in the python package \texttt{gwcosmo} version 1.0.0, can be found in \cite{Gray2020}.
One crucial step in the aforementioned method is to determine the probability that the host galaxy of an event is contained within the galaxy catalogue used for analysis. This is dependent on, amongst other factors, the completeness of the galaxy catalogue. It is also important not to introduce selection effects in assigning probabilities to potential host galaxies. Because of this, observed galaxies that are fainter than the apparent magnitude threshold adopted are discarded in the analysis. This avoids introducing a bias towards brighter galaxies as potential hosts in regions of the catalogue where fainter galaxies are not observed due to the flux-limited nature of the surveys.\footnote{Galaxy catalogue incompleteness can also impact adversely on gravitational-wave parameter estimation with imprecisely localised events. See e.g. \cite{mo2024usegalaxycatalogsgravitationalwave}.} A careful analysis to estimate the magnitude threshold for any sample of galaxies is therefore necessary.

GLADE and GLADE+ \citep{10.1093/mnras/sty1703,10.1093/mnras/stac1443} are composite catalogues made up of several surveys of varying depth and coverage. The GLADE+ catalogue comprises the GWGC, 2MPZ, 2MASS XSC, HyperLEDA and WISExSCOSPZ galaxy catalogues, and contains 22.5 million galaxies. Also included is the SDSS-DR16Q quasar catalogue. The GLADE+ catalogue is complete up to a luminosity distance of $d_{L} = 47^{+4}_{-2}$\,Mpc and contains bright galaxies up to 90\% of the total expected $B$-band and $K$-band luminosities up to $\sim$130\,Mpc \citep{10.1093/mnras/stac1443}. The previous GLADE catalogue does not contain the WISExSCOSPZ galaxy survey, and is complete to $d_{L} = 37^{+3}_{-4}$\,Mpc of the cumulative $B$-band galaxy luminosity. It contains $\sim$3 million objects that are categorised as galaxies \citep{10.1093/mnras/sty1703}.

Determining the completeness of a galaxy catalogue is not straightforward. In the context of GW cosmology, catalogue incompleteness can be treated in a number of ways \citep{Palmese_2023, Dalang:2023ehp}. In this paper we follow the approach adopted by \texttt{gwcosmo} and define the completeness of a galaxy catalogue as an inherent characteristic of a flux-limited survey, which can be modelled through the identification of a limiting apparent magnitude to which the catalogue is considered complete. The flux-limited nature of a galaxy catalogue will mean that fainter galaxies are only observable nearby, while brighter galaxies will be missing from the survey at increasing distances.

The current implementation of \texttt{gwcosmo} uses the median apparent magnitude of the galaxy sample to define the limiting apparent magnitude in the galaxy catalogue method. While this conservative estimate seeks to avoid any biases due to selection effects, it also discards information about the potential host galaxies of dark sirens.   In this paper, we describe the implementation within \texttt{gwcosmo} 1.0.0 of a robust test for determining the limiting apparent magnitude of a galaxy catalogue. In the main analysis, the test is applied to the {\em pixelated\/} version of the pipeline, as described in \cite{10.1093/mnras/stac366}. This version divides galaxy catalogues into HEALPix pixels which are treated separately in the analysis before being recombined in the final posterior. The pixelated implementation of \texttt{gwcosmo} introduced the treatment of galaxy catalogue completeness as directionally-dependent. The galaxy magnitude-redshift samples considered in this work are determined by the pixels used by \texttt{gwcosmo}.

The remainder of this paper is organised as follows. Section \ref{sec:methods} describes the robust test and its implementation into \texttt{gwcosmo}. Section \ref{sec:toymodel} then first applies the robust test to a toy model adapted from the analysis outlined in \cite{hitchhiker}, to illustrate some basic features of magnitude incompleteness on inferring the Hubble constant and how the application of the robust test mitigates these effects. Section \ref{sec:testcase} then describes further test cases demonstrating our methodology using \texttt{gwcosmo} with real GW data and a simulated galaxy catalogue. Section \ref{sec:results} next presents results of the new analysis using real data from the GWTC-1 and GWTC-3 catalogues. Finally, section \ref{sec:conc} presents a discussion of the results and future work to be carried out. Supplementary material illustrating the impact of some of the choices made in applying the robust test can be found in Appendix \ref{sec:appendix}.

\section{Methods}
\label{sec:methods}

\subsection{The statistical test}
\label{sec:robust}
We apply the statistical test for determining galaxy catalogue completeness first outlined in \cite{Rauzy2001}, hereafter R01. We will henceforth refer to this test as the robust method for estimating galaxy catalogue completeness. The robust method allows for the rigorous inference of the limiting apparent magnitude of a redshift-magnitude sample of galaxies.

We present an overview of the method here, while a full derivation can be found in R01. The test is related to a statistical test derived in \cite{1992ApJ...399..345E}, hereafter EP92.  By comparing two samples from the galaxy catalogue itself, rather than comparing a sample to the expected number of galaxies, this method assumes no specific model for the luminosity function, though it is assumed that the luminosity function is the same everywhere.

The method defines a statistic $T_{C}$ that tests different limiting apparent magnitudes. Essentially, the statistic tests whether or not the catalogue is emptier than expected for a given `trial' limiting apparent magnitude $m_{\rm lim}$.

For each sample of galaxies $\left \{ (m_{i},z_{i}) \right \}\rm $, we first define an array of test magnitude thresholds $m_{\rm lim}$, to which we will apply the statistical test. Here we obtain magnitude and redshift samples of galaxies from the "pixels" delineated by \texttt{gwcosmo}. We also use the GLADE, GLADE+ and DES catalogues as used by previous cosmological analyses. The apparent magnitudes for the galaxies in these catalogues are corrected for galactic extinction.

For each trial magnitude threshold $m_{\rm lim}$ we can define, for a galaxy in the catalogue with distance modulus $\mu$, a corresponding absolute magnitude $M_{\rm lim}(\mu)$ such that:

\begin{equation}
M_{\rm lim}(\mu) = m_{\rm lim} - \mu\text{.}
\label{eq:Mlim}
\end{equation}

Here $\mu$ can be computed as a function of the redshift of the galaxy assuming a fiducial cosmological model.  For example, we can follow \cite{Gray2020} and assume for simplicity that the redshift $z$ and luminosity distance $d_L$ of the galaxy are related by a linear Hubble law with a fiducial value of the Hubble constant, $H_0^*$, so that the distance modulus of the galaxy reduces to
\begin{equation}
\mu = 5 \log_{10} cz - 5 \log_{10} H_0^* + 25 ,
\label{eq:dmod}
\end{equation}
where $c$ is the speed of light. In this way eqs. \ref{eq:Mlim} and \ref{eq:dmod} define, for the adopted fiducial value of the Hubble constant, $H_0^*$, the faintest absolute magnitude, $M_{\rm lim}(\mu)$, at which a galaxy with distance modulus $\mu$ would be visible in the sample.

For the $i^{\rm th}$ galaxy with $(M_{i}, \mu_{i})$ we can define a random variable $\zeta_{i}$, defined as:
\begin{equation}
\label{eq:zeta}
\zeta_{i} = \frac{F(M_{i})}{F(M_{\rm lim}(\mu_{i}))} ,
\end{equation}
where $F$ is the cumulative luminosity function. Under the null hypothesis that the luminosity function is the same everywhere, and providing the trial apparent magnitude limit is not fainter than the {\em true\/} apparent magnitude limit of the catalogue, the variable $\zeta$ is distributed uniformly over the interval $[0,1]$. Moreover, it is also shown in R01 that $\zeta$ and $\mu$ are independent. 

R01 goes on to show that $F$, and hence $\zeta$, can be estimated simply by counting the number of galaxies that are found in certain regions of the $\left ( M,\mu \right )$ plane.   Specifically, for the $i^{\rm th}$ galaxy with absolute magnitude $M_{i}$, R01 defines the following two regions:
\begin{trivlist}
    \item{$S_1$: the region defined by $\mu < \mu_{i}$ and $M < M_{i}$}
    \item{$S_2$: the region defined by $\mu > \mu_{i}$ and $M_{\rm lim} (\mu_{i}) > M > M_{i}$}
\end{trivlist}

R01 then shows (following EP92) that $\zeta_i$ can be estimated as: 
\begin{equation}
    \zeta_{i} = \frac{r_{i}}{n_{i}+1} \text{,}
\end{equation}
where $r_i$ and $n_i$ are equal to the number of galaxies in the regions $S_1$ and $S_1 \cup S_2$ respectively. 

\begin{figure}
    \centering
    \includegraphics[width=.5\textwidth]{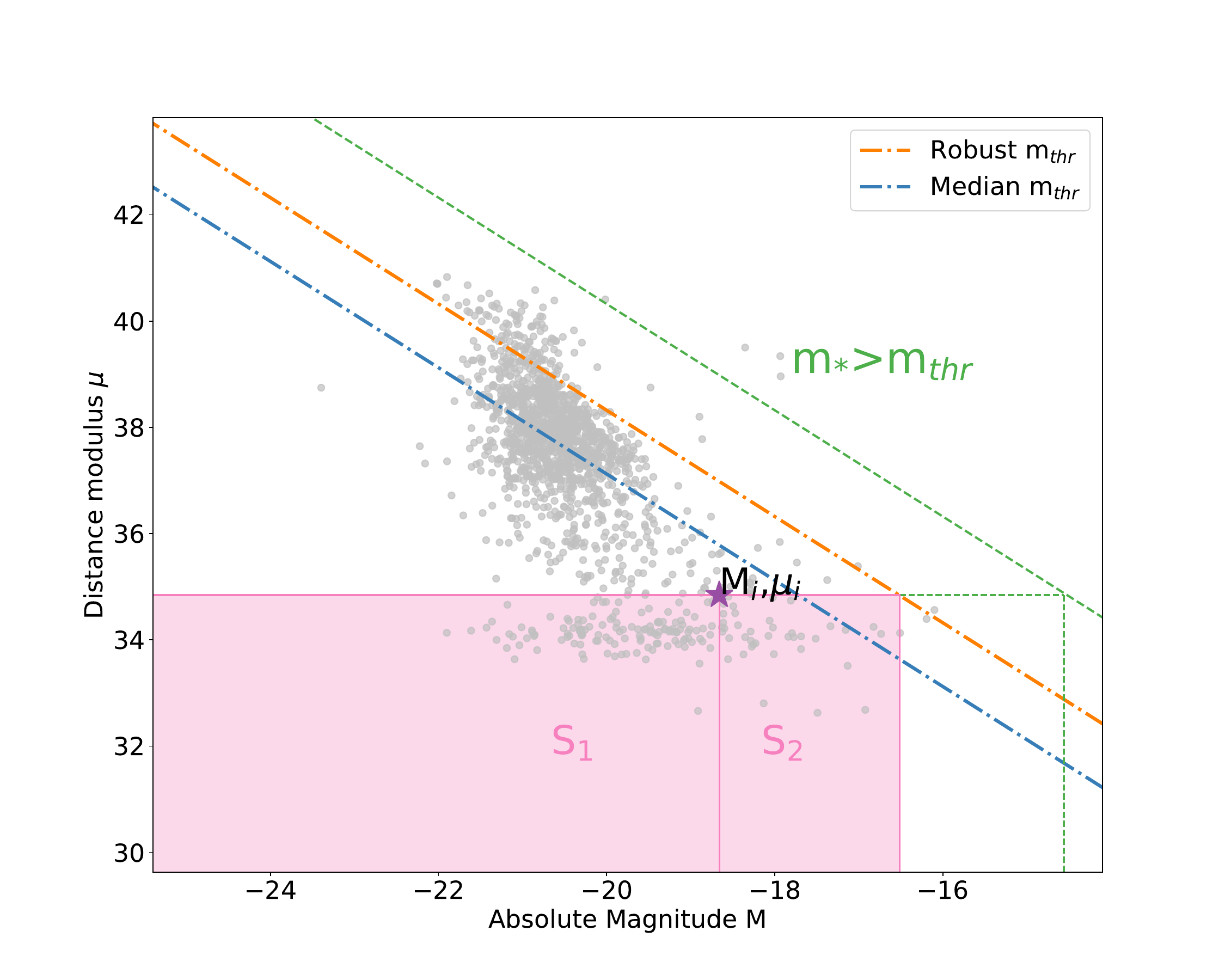}
    \caption{Illustrating the $S_1$ and $S_2$ areas for a single pixel in the GLADE $B$-band. The figure plots absolute magnitude $M$ vs distance modulus $\mu$ and shows how $S_1$ and $S_2$ are constructed for the $i^{\rm th}$ galaxy with values ($M_{i}$,$\mu_{i}$). The dash-dotted orange and blue lines show, respectively, the robust and median apparent magnitude thresholds for this sample of galaxies. The green line shows a trial limit magnitude $m_{*}$ that is fainter than the true $m_{\rm thr}$.}
    \label{fig:MZ_onepixel}
\end{figure}

Figure \ref{fig:MZ_onepixel} illustrates how the areas $S_1$ and $S_2$ are constructed, for a galaxy in one pixel of the $B$-band of GLADE+. Also plotted on the figure are the $(M,\mu)$ values of the other galaxies within that pixel. In each subsample $S_1$ and $S_2$, $M$ and $\mu$ are, by construction, independent. 

It is important to note that the choice of fiducial Hubble constant for computing $M_{\rm lim}(\mu)$ in Eq. \ref{eq:Mlim} is arbitrary and has no effect on the observed values of $r_i$ and $n_i$, nor on the computed values of $\zeta_i$. Choosing a different fiducial Hubble constant would shift the positions of the $(M,\mu)$ values in Fig. \ref{fig:MZ_onepixel}, but would also shift in exactly the same way the boundaries of the regions $S_1$ and $S_2$ defined for the $i^{\rm th}$ galaxy, so that the number count of galaxies in each of these regions is unchanged.

As follows from EP92, the expected value of $r_{i}$ is $\frac{1}{2} (n_{i}+1)$ and the expected value $E_{i}$ and variance $V_{i}$ of $\zeta_{i}$ are:

\begin{equation}
  E_{i} = \frac{1}{2}  \text{    and    }  V_{i} = \frac{n_{i}-1}{12(n_{i}+1)}\text{.}
\end{equation}

R01 then defines the quantity $T_{C}$ as:
\begin{equation}
    T_{C} = \frac{\sum\limits^{N}_{i=1} \Big(\zeta_{i}-\frac{1}{2}\Big)}{\sum\limits^{N}_{i=1}V_{i}} ,
\end{equation}
and goes on to show that, under the null hypothesis, the expectation of $T_{C}$ is zero and its variance is unity. Moreover, $T_{C}$ can be estimated without any a-priori assumptions about the form of the luminosity function. If the sample is complete to the trial value $m_{\rm lim}$, then $T_{C}$ follows a Gaussian distribution around zero with unit variance.  However, $T_{C}$ starts going systematically negative when there is a deficit of galaxies fainter than $M_{\rm lim}(\mu_i)$, for the $i^{\rm th}$ galaxy. This deficit will occur when the trial apparent magnitude limit, $m_{\rm lim}$, is fainter than the {\em true\/} magnitude threshold $m_{\rm thr}$.

The statistic $T_{C}$ becoming systematically very negative is therefore indicative of catalogue incompleteness. This is a result of the region $S_2$ becoming systematically emptier due to the impact of the apparent magnitude limit. Following \cite{Rauzy2001}, we take $T_{C}=-3$ as a threshold value that represents the limiting magnitude for catalogue completeness.

\subsection{Implementation}

The robust statistical test summarised in Section \ref{sec:robust} is implemented using the pixelated version of \texttt{gwcosmo}. A full overview of the pixelated method can be found in \cite{10.1093/mnras/stac366}. In this method, galaxy catalogues are divided into directional pixels of equal area, with each pixel containing a magnitude-redshift sample of galaxies that lie within a given range of right ascension and declination. \texttt{gwcosmo} requires the adoption of a limiting apparent magnitude threshold, $m_{\rm thr}$, for that pixel; in \cite{10.1093/mnras/stac366} $m_{\rm thr}$ is taken to be the median apparent magnitude in the pixel, whereas in our work the median is replaced by the magnitude limit determined by the robust method applied to the galaxies in the pixel.  

The value of the magnitude threshold affects in two ways the resulting posterior on $H_{0}$ that is computed by \textbf{\texttt{gwcosmo}}. Firstly, all galaxies fainter than $m_{\rm thr}$ are discarded from the analysis, so approximating $m_{\rm thr}$ to be the median apparent magnitude means that half of the catalogue is discarded. Secondly, the threshold affects the calculated probability that the host galaxy of the GW event is within the galaxy catalogue.

In order to compute a final posterior on $H_{0}$, \texttt{gwcosmo} marginalises the probability of the GW data $x_{\rm GW}$ over the two propositions $G$ and $\overline{G}$, where:
\begin{itemize}
\item $G$ denotes the proposition that the host galaxy is within the galaxy catalogue,
\item $\overline{G}$ denotes the proposition that the host galaxy is outwith the galaxy catalogue.
\end{itemize}

Each of these propositions depends on the apparent magnitude threshold of the catalogue. For example, to evaluate the probability that the proposition $\overline{G}$ is true involves integrating over the portion of the galaxy luminosity function that cannot be observed due to being fainter than the catalogue's limiting apparent magnitude.

To unpack this in more detail, and following \cite{Gray2020} Appendix 2, the conditional probability of obtaining a GW signal with data $x_{\rm GW}$, given the signal's detection $D_{\rm GW}$ and a Hubble constant $H_{0}$, may be written as the sum, over the propositions $g \in \{ G, \overline{G}\}$, of the product of the conditional probability of $x_{\rm GW}$ given $g$ and the conditional probability of $g$ given $D_{\rm GW}$ and $H_{0}$, i.e.

\begin{equation}
p(x_{\rm GW}|D,H_{0}) = \sum_{g \in \{ G, \overline{G}\}} p(x_{\rm GW}|g,D_{\rm GW},H_{0})p(g|D,H_{0})\text{.}
\label{eq:probx}
\end{equation}

Hence, the right hand side of eq. \ref{eq:probx} is made up of four terms that must be computed by \texttt{gwcosmo}: the two conditional probabilities, each of which is evaluated for both the $G$ and $\overline{G}$ cases. All four of these terms depend on the apparent magnitude threshold.

In the $B$-band, the GLADE and GLADE+ catalogues contain millions of galaxies, and each pixel contains up to thousands of galaxies at the typical resolution of $N_{\rm side} =32$ used in the analyses. The complexity of the robust test described above goes up with $\mathcal{N}_{\rm gal}^{2}$, where $\mathcal{N}_{\rm gal}$ is the number of galaxies in the pixel, making it computationally expensive and slow to determine the apparent magnitude threshold of each pixel for each gravitational wave event.
To overcome this, for analyses using the $B$-band, we randomly sample a subset of $\mathcal{N}_{\rm gal}=400$ galaxies with $(m_{i},z_{i})$ multiple times and evaluate $m_{\rm thr}$ for each subset of galaxies. The final value of the magnitude threshold $m_{\rm thr}$ for that pixel is then taken to be the mean of the evaluated thresholds for the different random sub-samples\footnote{Future instances of \texttt{gwcosmo} will compute the value of $m_{\rm thr}$ only once for each pixel across an entire galaxy catalogue rather than for each GW event, meaning that in the future this sub-sampling will not be necessary.}. This sub-sampling currently only needs to be carried out for the $B$-band of the GLADE and GLADE+ catalogues; in the $K$-band, each pixel is more sparsely populated. Example calculations for $m_{\rm thr}$ in one pixel at $N_{\rm side} = 32$, for different $\mathcal{N}_{\rm gal}$ in the MICE catalogue described in sections~\ref{sec:toymodel} and~\ref{sec:testcase}, are shown in figure~\ref{fig:Ngal_ms}.

Following R01, we use the threshold $T_{C} = -3$ to determine the apparent magnitude threshold. Different thresholds will affect the resulting $m_{\rm thr}$ and this is illustrated in figure~\ref{fig:TCs}. Since the value of $T_{C}$, especially in composite galaxy catalogues such as GLADE and GLADE+, can be noisy, taking $T_{C} = -3$ helps avoid underestimating $m_{\rm th}$ due to noisy data, while adopting larger (i.e. more negative) values of $T_{C}$ could lead to overestimated apparent magnitude thresholds.\footnote{A further investigation of the impact of both $\mathcal{N}_{\rm gal}$ and the threshold on $T_{C}$ on the main pipeline can be found in figures~\ref{fig:TCs_GW150914} and~\ref{fig:Ngal_GW150914} in appendix~\ref{sec:appendix}.}

The redshifts $z_{i}$ of the galaxies in the sample have an associated photometric or spectroscopic uncertainty $\sigma_{z_{i}}$. To propagate redshift uncertainties to the magnitude threshold estimate, we assume the uncertainties on the galaxy redshifts are described by a truncated Gaussian --- i.e., for each redshift $z_{i}$, the distribution that is sampled from is truncated, as appropriate, to only allow positive redshift values. 

In order to sample over the redshift distributions, the magnitude threshold $m_{\rm thr}$ is estimated for several samples of $(m_{i},z_{i})$ for each pixel. This in turn generates a range of $m_{\rm thr}$ values. Note that the apparent magnitudes $m_{i}$ also have an associated uncertainty $\sigma_{m_{i}}$. However, for direct comparison to the median method for estimating incompleteness, which does not incorporate $m_{i}$ uncertainties, we choose to ignore them in this work.

The trial magnitude thresholds $m_{\rm lim}$ for each sample are taken to lie between $m_{\rm lim} = m_{\rm med}-0.5$ and $m_{\rm lim} = m_{\rm med}+4$ with a step of $0.05$ mag, where $m_{\rm med}$ is median apparent magnitude for each galaxy-redshift sample. This was found to be the best range of test limiting apparent magnitudes that gave appropriate resolution without sacrificing efficiency. Future implementations to \texttt{gwcosmo} 2.0. will be able to pre-compute apparent magnitude thresholds for the entire galaxy catalogue, bypassing some of the concerns around computational efficiency and allowing the use of finer grids where desirable.

\section{Testing the robust method on a toy model analysis}\label{sec:toymodel}

In this section we carry out the analysis described above on a toy model based on the methodology outlined in \cite{hitchhiker}. This methodology is modified and extended to account for incompleteness, while ignoring localisation effects and luminosity weighting.

When assuming the catalogue is complete, the \cite{hitchhiker} analysis can be summarised by the following key equation:
\begin{equation}
p(x_{\rm GW}|D_{\rm GW}, H_{0}) = \frac{\sum^{N_{\rm gal}}_{i=1} \int p\big(x_{\rm GW}|d_{L}(z_{i}, H_{0})\big)p(z_{i})dz_{i}}{\sum^{N_{\rm gal}}_{i=1} \int p\big(D_{\rm GW}|d_{L}(z_{i}, H_{0})\big)p(z_{i})dz_{i}},
\label{eq:GairMain}
\end{equation}
where $p(z_{i})$ is the probability distribution for each redshift $z_{i}$ that is in the galaxy catalogue.

We now introduce incompleteness to the above equation, splitting the relevant term into a in-catalogue part and an outwith catalogue part, as is done in \texttt{gwcosmo}. As a result, Equation~\ref{eq:GairMain} is replaced by equation~\ref{eq:probx}. Following~\cite{hitchhiker}, we ignore the merger rate term for this toy model. The probability of the event having a host galaxy contained within the galaxy catalogue is then:
\begin{equation}
\label{Eq:G_DH0_end}
\begin{aligned}
\\ p(G|D_{\text{GW}},H_0)&= \dfrac{\int^{z_{\rm lim}}_0 dz\int dM p(D_{\text{GW}}|z,H_0)p(z)p(M|H_0)}{\int dz \int dM p(D_{\text{GW}}|z,H_0)p(z)p(M|H_0)},
\end{aligned}
\end{equation}
and the reciprocal probability that the host galaxy is outwith the catalogue:
\begin{equation}
p(\overline{G}|D_{\rm GW}, H_{0}) = 1 - p(G|D_{\rm GW}, H_{0}),
\end{equation}
with $z_{\rm lim} = z(m_{\rm thr}, M, H_{0})$, and $p(z)$ the redshift prior, here taken to be uniform in comoving volume.

The term for the probability of the data $x_{\rm GW}$ given that the host galaxy is outwith the galaxy catalogue, $p(x_{\rm GW}|\overline{G},D_{\rm GW},H_{0})$, is also introduced:

\begin{equation}
p(x_{\rm GW}|\overline{G},D_{\rm GW},H_{0}) = \frac{\int_{z_{\rm lim}}^{z_{\rm max}}dz\int dMp(x_{\rm GW}|z,H_{0})p(z)p(M|H_{0})}{\int_{z_{\rm lim}}^{z_{\rm max}}dz\int dM p(D_{\rm GW}|z,H_{0})p(z)p(M|H_{0})} .
\end{equation}

This toy model is applied using a simulated galaxy catalogue, MICECAT \citep{TALLADA2020100391, 2017ehep.confE.488C,10.1093/mnras/stv138,10.1093/mnras/stv1708,10.1093/mnras/stu2464,10.1093/mnras/stu2402,10.1093/mnras/stu2492}. MICECAT v2.0 is a simulated galaxy catalogue covering one octant of the sky. The catalogue takes the following cosmological parameters as inputs: $\Omega_{M}= 0.25$, $\sigma_{8} = 0.8$, $n_{s} = 0.95$, $\Omega_{b} = 0.044$, $\Omega_{\Lambda}=0.75$ and $h = 0.7$. In this analysis, we use the SDSS-like data generated for the $r$-band. For the sample of galaxies studied, the robust apparent magnitude limit is $m_{\rm thr} = 24.78$ and the median apparent magnitude is $m_{\rm thr} = 23.78$. 

\begin{figure}
\centering
\includegraphics[width=.5\textwidth]{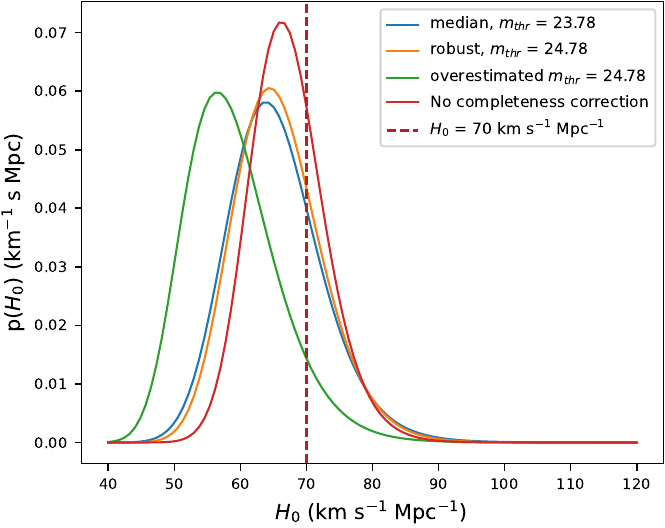} 
    \caption{Results from the toy model analysis with 300 events with no completeness correction (shown in red), with the robust magnitude threshold applied (orange), and with the median magnitude threshold applied (blue), for 300 events. The green curve shows results when we apply a magnitude cut at $m_{\rm thr} = 23$, but vastly overestimate the magnitude threshold in the analysis as $m_{\rm thr} = 24.78$.}
    \label{fig:toymodel_results}
\end{figure}

\begin{figure}
\centering
\includegraphics[width=.49\textwidth]{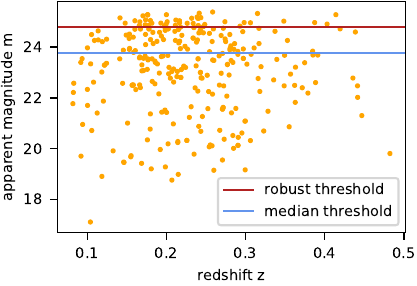} 
    \caption{Redshifts and apparent magnitudes of the host galaxies for the 300 events considered in the toy model analysis of Section \ref{sec:toymodel}. The red line is the robust apparent magnitude threshold, $m_{\rm thr} = 24.78$, while the blue line is the median apparent magnitude $m_{\rm thr} = 23.78$.}
    \label{fig:toymodel_events}
\end{figure}

As in~\cite{hitchhiker}, the analysis is run on a patch of sky, with no additional directional information used; specifically, we consider a patch of sky of area 0.1 square radians containing 11070 galaxies. Figure~\ref{fig:toymodel_results} shows results of the analysis for $\mathcal{N}_{\rm GW} = 300$ GW events with $\sigma_{d_{L}}/d_{L} = 0.2$. We considered the following scenarii: using the median apparent magnitude threshold, using the robust apparent magnitude threshold, the original analysis with no completeness correction required (i.e. we assume a complete galaxy survey) and lastly a case where we apply an apparent magnitude cut to the catalogue at $m_{\rm thr} = 23$ but erroneously assume a threshold of $m_{\rm thr} = 24.78$. We can see from Figure~\ref{fig:toymodel_results} that the last scenario results in a posterior biased towards a lower $H_{0}$. We can also see that, as expected, the posterior on $H_{0}$ for the robust case shows a slight improvement over that for the median case. Figure~\ref{fig:toymodel_events} shows the redshift and apparent magnitude of each event considered in our analysis, along with the robust and median apparent magnitude thresholds. Of these 300 GW events, 253 lie within galaxies whose apparent magnitude is brighter than the robust magnitude threshold. However, of these 253 events, 79 are from host galaxies that are fainter than the median apparent magnitude. 

\begin{figure}
\centering
\includegraphics[width=.5\textwidth]{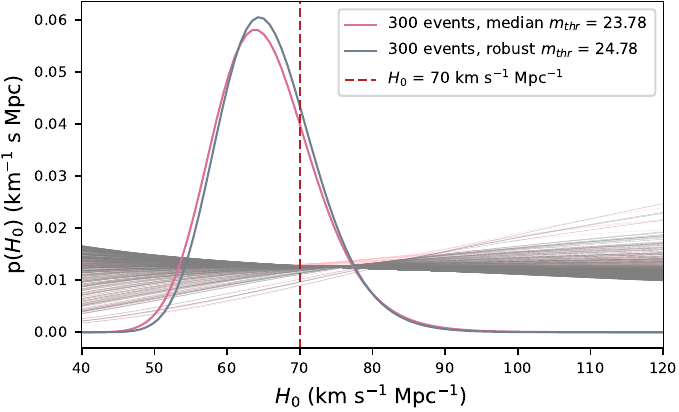} 
    \caption{Posteriors on $H_{0}$ for each of the 300 events in the analysis, along with the combined posteriors for the robust and median methods.}
    \label{fig:toymodel_allevents}
\end{figure}

\begin{figure}
\centering
\includegraphics[width=.5\textwidth]{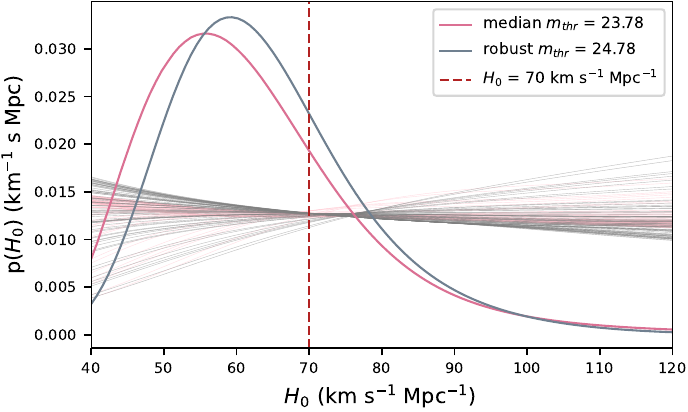} 
    \caption{Posterior on $H_{0}$ for those 79 events with a host galaxy apparent magnitude in-between the median and robust magnitude thresholds, and their corresponding combined posterior.}
    \label{fig:toymodel_inbetweenevents}
\end{figure}

Figure~\ref{fig:toymodel_allevents} shows the individual and combined posteriors on $H_{0}$ for both methods with all 300 events, while Figure~\ref{fig:toymodel_inbetweenevents} shows the posteriors on $H_{0}$ when only considering those 79 events with host galaxies that lie in-between the robust and median apparent magnitude thresholds. The 68\% confidence interval using all 300 GW events gives $H_{0}=64.4^{+7.1}_{-6.2}{\text{\,km\,s}}^{-1}{\text{\,Mpc}}^{-1}$ when using the robust method, and $H_{0}=63.8^{+7.7}_{-6.2}{\text{\,km\,s}}^{-1}{\text{\,Mpc}}^{-1}$ when using the median apparent magnitude threshold. When only the 79 events in-between the two values for the apparent magnitude threshold are considered, using the robust method gives $H_{0}=59.2^{+14.0}_{-10.4}{\text{\,km\,s}}^{-1}{\text{\,Mpc}}^{-1}$, while using the median apparent magnitude as a threshold yields $H_{0}=55.7^{+14.4}_{-11.9}{\text{\,km\,s}}^{-1}{\text{\,Mpc}}^{-1}$.

\section{\texttt{gwcosmo} Test cases using a simulated galaxy catalogue}
\label{sec:testcase}

In this section we further demonstrate the impact of a fainter magnitude limit using a simulated MICE galaxy catalogue. We use the MICECAT v2.0 galaxy catalogue to populate the rest of the sky for application with \texttt{gwcosmo}. As with the toy model considered in the previous section, we use the SDSS-like data generated for the $r$-band. We choose the $r$-band as it is well described by a Schechter function with parameters $\Phi^{*} = (1.49 \pm 0.04 ) \cdot 10^{-2}h^{3} \rm Mpc^{-3}$, $M^{*}-5\log_{10}h = -20.44 \pm 0.01$ and $\alpha = -1.05 \pm 0.01$ \citep{10.1093/mnras/stu2402, Blanton_2003}.

\begin{figure*}
\centering
\includegraphics[width=.49\textwidth]{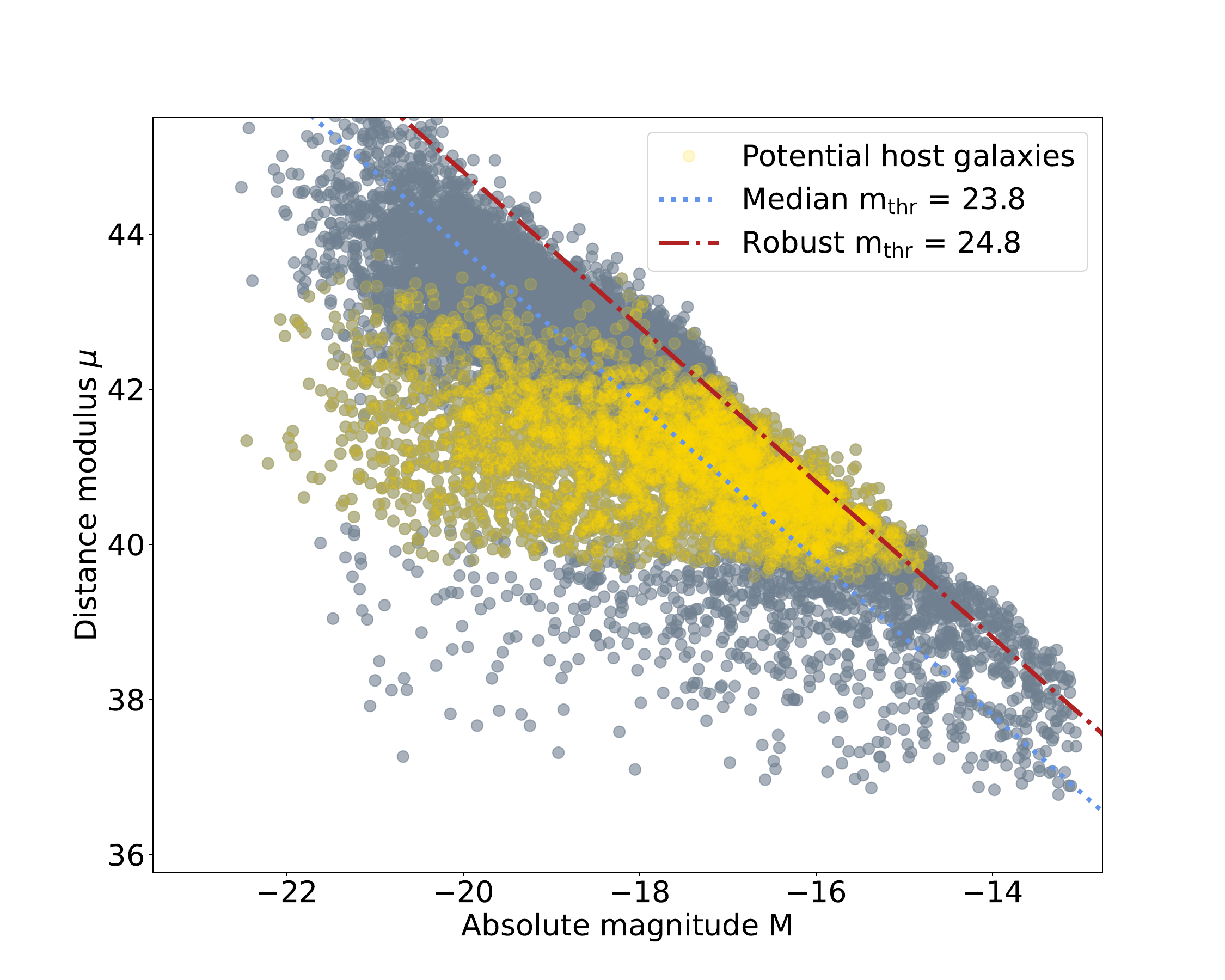} \includegraphics[width=.49\textwidth]{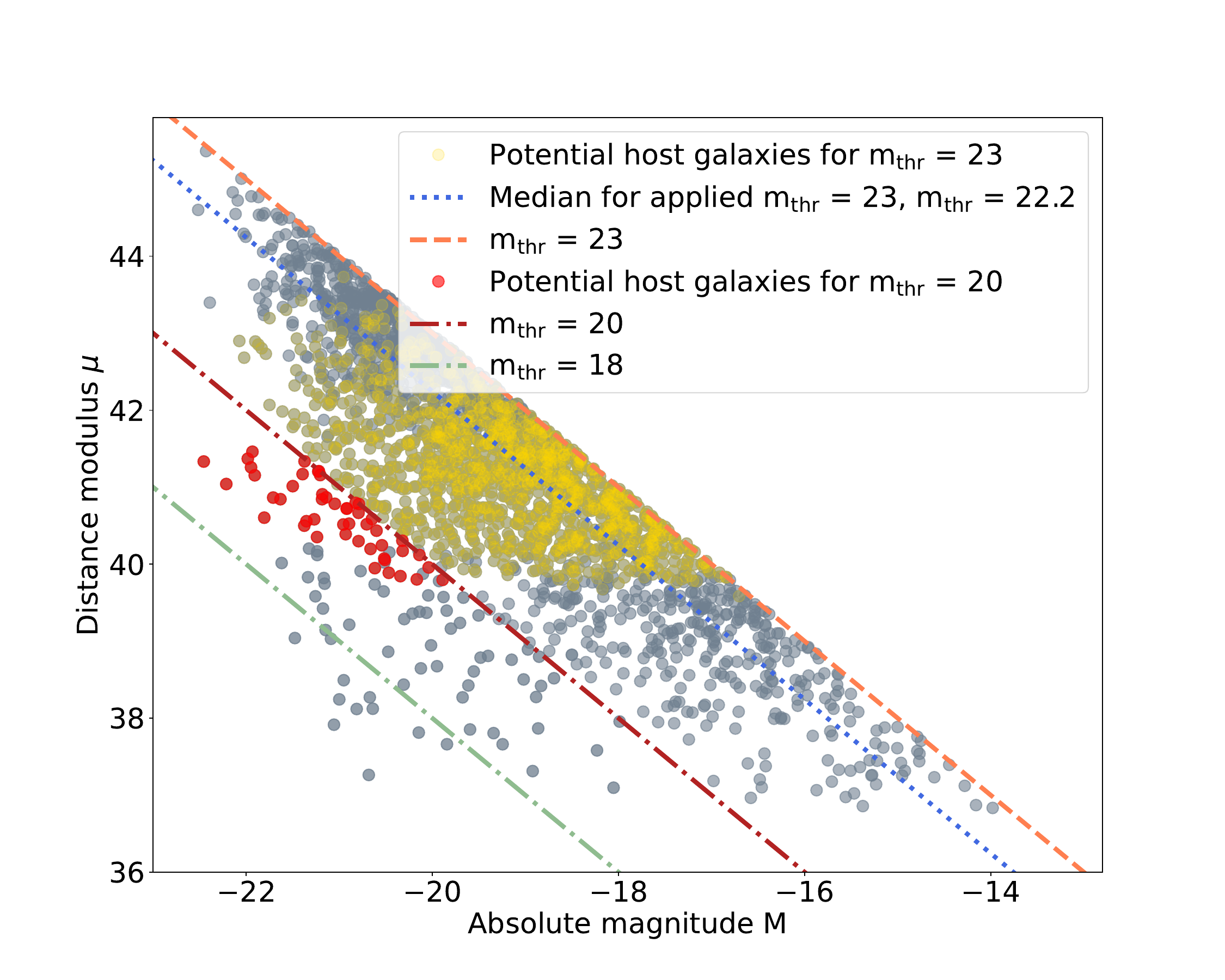}
    \caption{Left panel: Illustrating the robust and faint magnitude limits for a sub-sample of 10000 galaxies from the MICECAT v2.0 catalogue. The evolution-corrected SDSS-like $r$-band data is used. Galaxies in gold (i.e. potential host galaxies) are within the redshift boundaries associated with GW170729. For this sample, we find $m_{\rm thr, robust} = 24.8 $ and $m_{\rm thr, median} = 23.8$. Right panel: The same sub-sample of galaxies but now applying a true apparent magnitude threshold at $m_{\rm thr} = 23$, shown by the orange dashed line. In this case the median apparent magnitude limit, shown by the blue dotted line, lies at $m_{\rm lim} = 22.2$.  Adopting progressively brighter limiting magnitudes, however, severely restricts the number of potential host galaxies that remain in the sample.  Shown in red are the remaining potential host galaxies when $m_{\rm thr} = 20$, as indicated by the red dashed-dotted line. No potential hosts remain in the sample when $m_{\rm thr} = 18$, as indicated by the green dashed-dotted line.}
\label{fig:MICEcat}
\end{figure*}

The robust method is first applied to a sub-sample of 10000 galaxies from this simulated catalogue and the robust magnitude limit in the $r$-band is determined to be $m_{\rm thr} = 24.8$, while the median is $m_{\rm thr} = 23.8$.\footnote{Since the SDSS-like data has a bright magnitude limit along with its faint limit, the robust method in the form introduced in R01 has some limitations --- see the discussion in \cite{2007MNRAS.376.1757J}. However, the steep decline of the $T_C$ statistic still provides a clear signature of the faint magnitude limit in this case. In a future instance of \texttt{gwcosmo} the robust statistical test will be incorporated with both a bright and faint limit, following \cite{2007MNRAS.376.1757J}.}  The two panels of Figure \ref{fig:MICEcat} show the distribution of distance modulus versus absolute magnitude for the galaxies in this sub-sample. In the left panel the median and robust magnitude limits are shown as the (blue) dotted line and (red) dot-dash line respectively.

In order to connect this illustrative example to a real GW source, we consider the event GW170729: a massive and distant binary black hole system source at $d_{L} =2840^{+1400}_{-1360}$Mpc or equivalent $z = 0.49^{+0.19}_{-0.21}$\citep{TheLIGOScientificCollaboration2018GWTC-1:Runs}. The left panel of Fig. \ref{fig:MICEcat} shows in gold those galaxies that lie within the redshift boundaries associated with GW170729 --- i.e. these would be the potential host galaxies of this source if it had been observed in the simulated universe represented by our MICECAT sub-sample.
The right panel of Fig. \ref{fig:MICEcat} then illustrates how the adoption of different, progressively brighter, faint magnitude limits will severely affect the number of potential host galaxies in our analysis.

\begin{figure*}
\centering
\includegraphics[width=.49\textwidth]{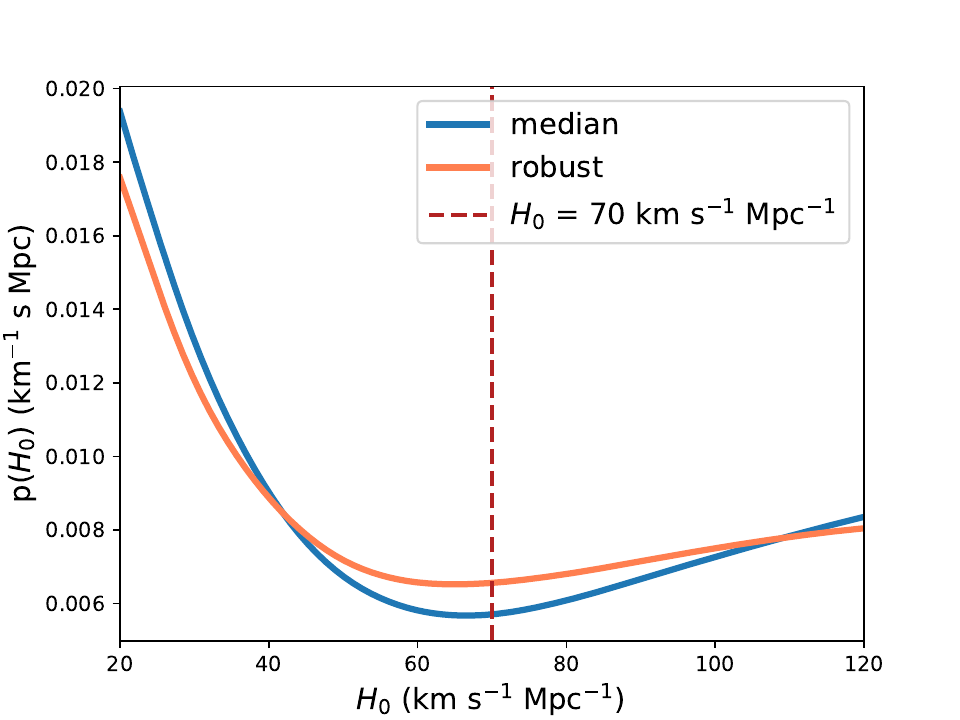} \includegraphics[width=.49\textwidth]{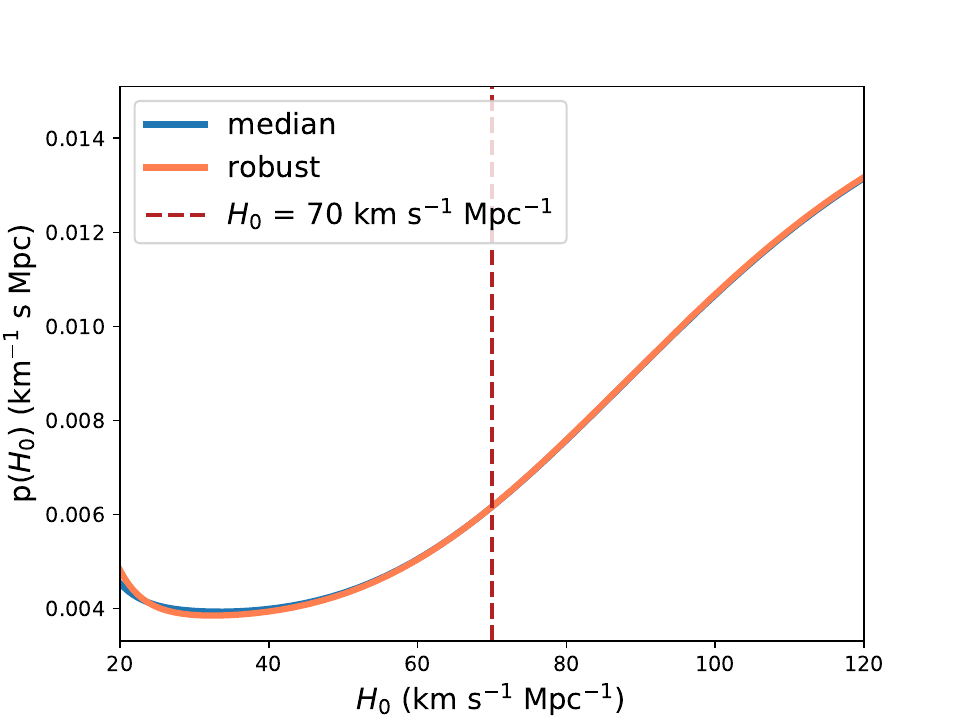}
    \caption{Posterior distributions, $p(H_{0})$, obtained for an illustrative mock GW event that combines GW170729 with our simulated MICECAT v2.0 galaxy catalogue. Results are shown for two imposed true values of the apparent magnitude threshold, at $m_{\rm thr} = 23$ (left panel) and $m_{\rm thr} = 18$ (right panel). When $m_{\rm thr} = 18$, the posteriors revert to the ``empty" catalogue case regardless of whether the median apparent magnitude or the robust method is used to estimate the magnitude threshold. On the other hand, with a true magnitude threshold of $m_{\rm thr} = 23$, while the catalogue is still very empty at the distances associated with GW170729, using the robust method to estimate the magnitude threshold still results in more support from the catalogue for values of $H_0$ close to the assumed value of $70 {\text{ km s}}^{-1}{\text{ Mpc}}^{-1}$.}
    \label{fig:GW170729}
\end{figure*}

Figure \ref{fig:GW170729} then shows the results of the \texttt{gwcosmo} analysis for GW170729, after imposing two different values of the true apparent magnitude threshold to the simulated galaxy catalogue: $m_{\rm thr} = 23$ (left panel) and $m_{\rm thr} = 18$ (right panel) respectively. Both panels show the posterior for $H_0$, assuming a true value of $70 {\text{ km s}}^{-1}{\text{ Mpc}}^{-1}$. Despite the catalogue not being complete at $m_{\rm thr} = 23$, there is still increased support for values of $H_0$ close to the true value when using the robust method to estimate the magnitude threshold, compared with using the median apparent magnitude.  When a true magnitude threshold of $m_{\rm thr} = 18$ is imposed, however, the $H_0$ posterior reverts to the "empty catalogue" case regardless of which method is used to estimate that threshold.

\begin{figure}
\centering
\includegraphics[width=.5\textwidth]{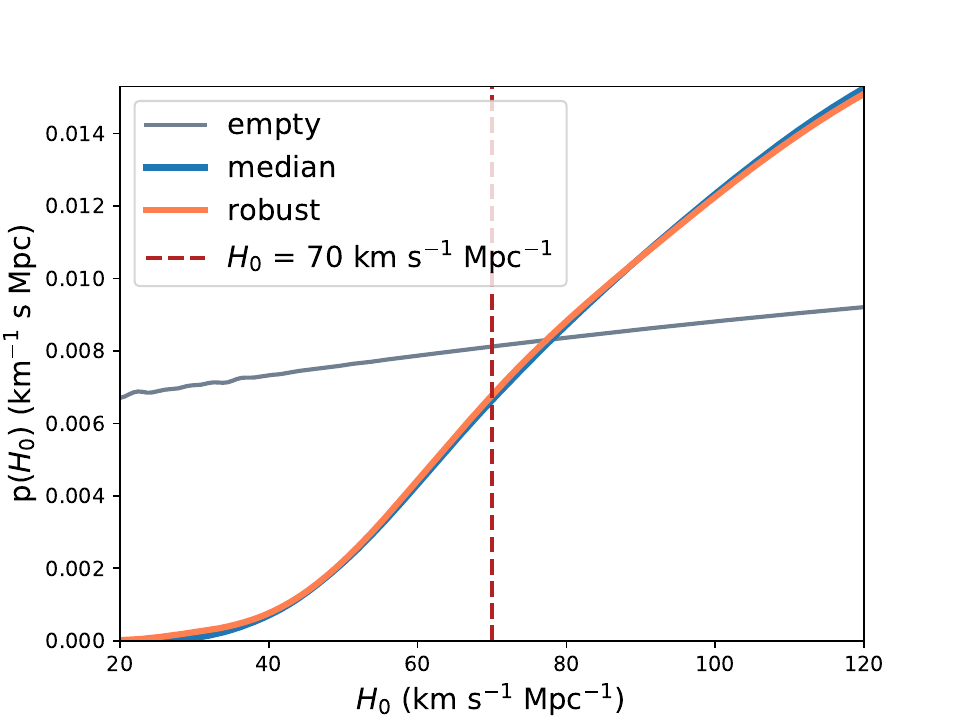}
    \caption{Posterior distributions, $p(H_{0})$, obtained for an illustrative mock GW event that combines GW170608 with our simulated MICECAT v2.0 galaxy catalogue. Results are shown for a median apparent magnitude $m_{\rm thr}=23.8$, robust $m_{\rm thr} = 24.8$  and the empty catalogue case. The event is highly informative regardless of the apparent magnitude threshold method used.}
    \label{fig:GW170608}
\end{figure}

Next, another test case is then carried out on the well-localised event GW170608, a binary black hole with a luminosity distance $d_{L} = 340^{+140}_{-140}$\,Mpc~\citep{GW170608}. Figure~\ref{fig:GW170608} shows results for a patch of sky analysis using the median apparent magnitude limit, robust apparent magnitude limit, and empty catalogue cases. Since this event is already highly informative when using the median apparent magnitude threshold, the improvement achieved by using the robust apparent magnitude threshold is negligible.

\section{Results}
\label{sec:results}
The previously outlined robust method of estimating the apparent magnitude threshold $m_{\rm thr}$ is now implemented within \texttt{gwcosmo} and applied to the inference of $H_{0}$ using the GWTC-1 \citep{Abbott_2021} and GWTC-3 \citep{Abbott_2023} catalogues of gravitational wave events. As was the case in the LVK analyses previously performed on those catalogues, only events with an SNR above $11$ are considered in this work.

\subsection{GWTC-1}
GWTC-1 is made up of all compact binary coalescences detected up to the end of the second observing run of the LVK network \citep{GWTC1}. Of the eleven events in GWTC-1, seven are used in the inference of $H_{0}$. Six of them are binary black holes, and one of them is the BNS event GW170817 which had an associated EM counterpart; it is not affected by any changes to the catalogue method.

The GWTC-1 data has previously been used in the inference of $H_{0}$ using the GLADE catalogue. Here we repeat the analysis using the $B$-band of both the GLADE and GLADE+ catalogues. One caveat to note is that the assumption that the luminosity function is universal might fail when using the $B$-band of the GLADE+ catalogue; therefore, this analysis should serve as proof of principle for consequent analyses.

\begin{figure}
    \centering
    \includegraphics[width=.5\textwidth]{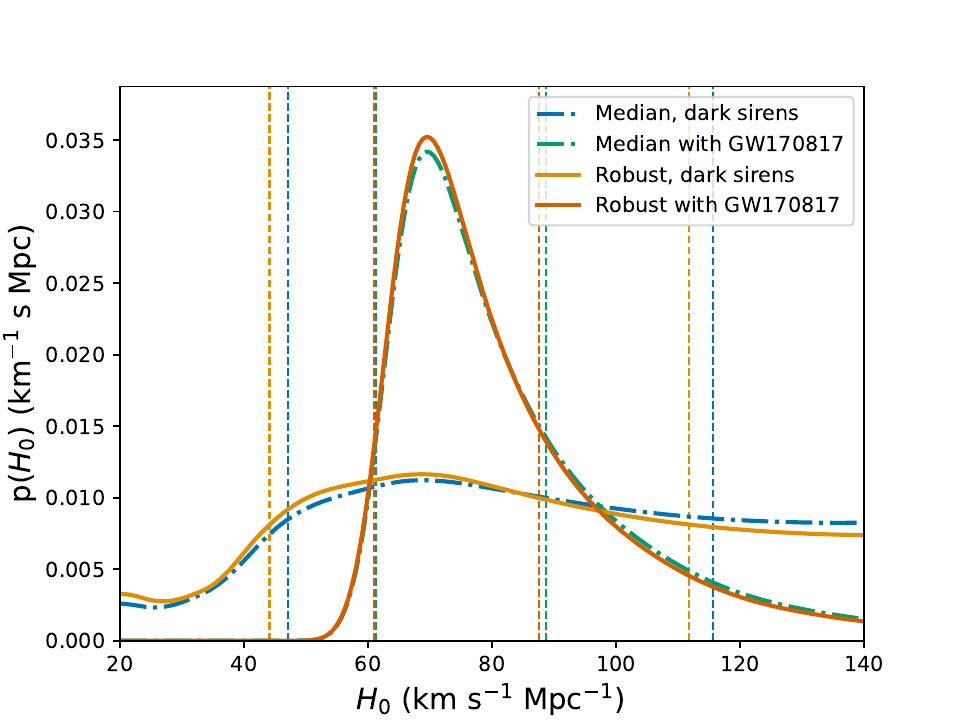}
    \caption{Final posterior on $H_{0}$ using the GLADE $B$-band on the GWTC-1 dataset. Vertical dashed lines show the $1\sigma$ intervals. The solid lines show results showing the robust method, while dash-dotted lines show results using $m_{\rm med}$ as the apparent magnitude threshold.}
    \label{fig:GWTC1}
\end{figure}

In the GWTC-1 analysis using the $B$-band of the GLADE and GLADE+ catalogues, there is a change in the recovered posterior on $H_{0}$ when using the robust method of estimating $m_{\rm thr}$ compared to the final posterior using $m_{\rm med}$ as the threshold. Figure \ref{fig:GWTC1} shows results for the GWTC-1 analysis using the GLADE $B$-band. The width of the final  recovered posterior on $H_{0}$ using the robust method is 1.3\% narrower than that of the median method for the 68.3\% percentile when considering only dark sirens. The final posterior with GW170817 is 3.4\% narrower using the robust method.

\begin{figure}
    \centering
    \includegraphics[width=.5\textwidth]{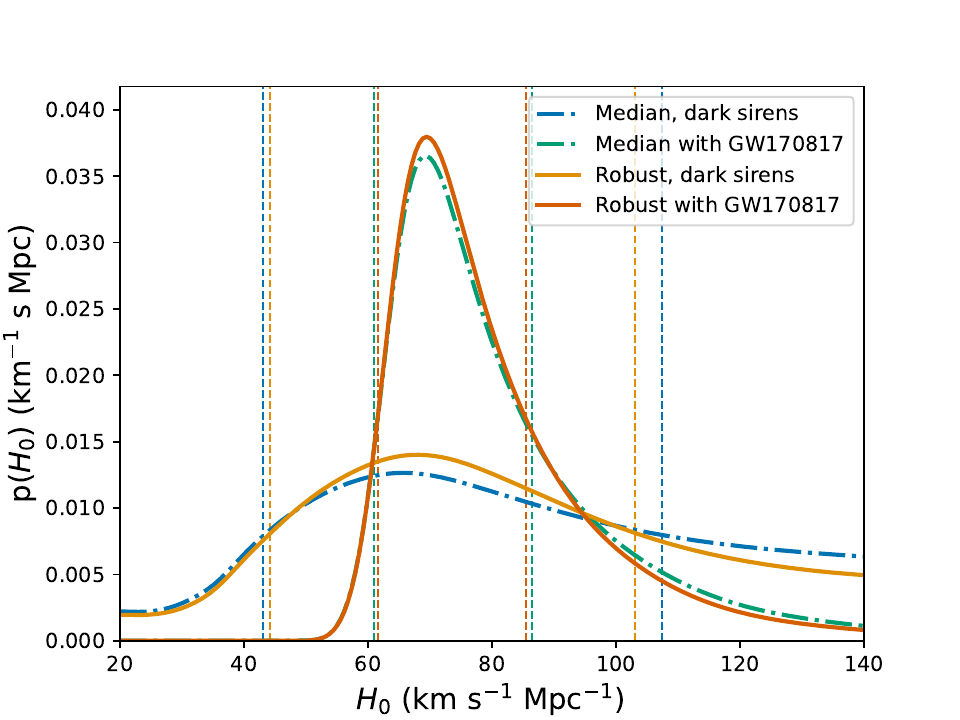}
    \caption{Final posterior on $H_{0}$ using the GLADE+ $B$-band on the GWTC-1 dataset. Vertical dashed lines show the $1\sigma$ intervals. The solid lines show results showing the robust method, while dash-dotted lines show results using $m_{\rm med}$ as the apparent magnitude threshold.}
    \label{fig:GWTC1+}
\end{figure}

Figure \ref{fig:GWTC1+} shows the same analysis using the GLADE+ $B$-band instead of the GLADE $B$-band. GW170814 remains unchanged, being analysed using the DES \citep{DESY1} catalogue. For this analysis, when using robust, the final posterior on $H_{0}$ is 8.6\% narrower when considering only dark sirens, and 6.3\% narrower when considering both dark and bright sirens. This is a clear improvement to the GWTC-1 results with GLADE+ when using the robust method.

\begin{figure}
    \centering
    \includegraphics[width=.5\textwidth]{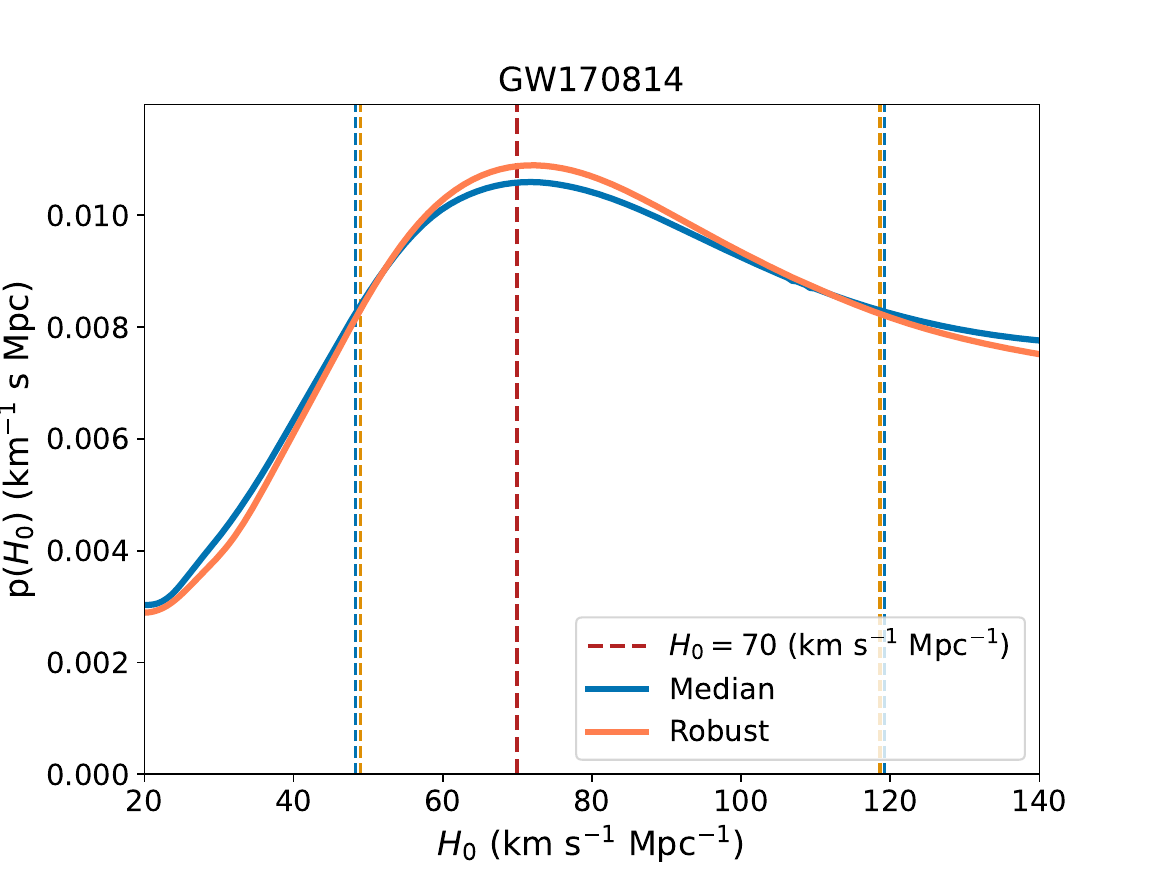}
    \caption{Posterior on $H_{0}$ for the event GW170814 using the DES catalogue. The blue line shows results using the median apparent magnitude as $m_{\rm thr}$, and the orange line shows the final posterior when the robust method of inferring $m_{\rm thr}$ is used.}
    \label{fig:GW170814_DES}
\end{figure}

\begin{figure*}
    \centering
    \includegraphics[width=.8\textwidth]{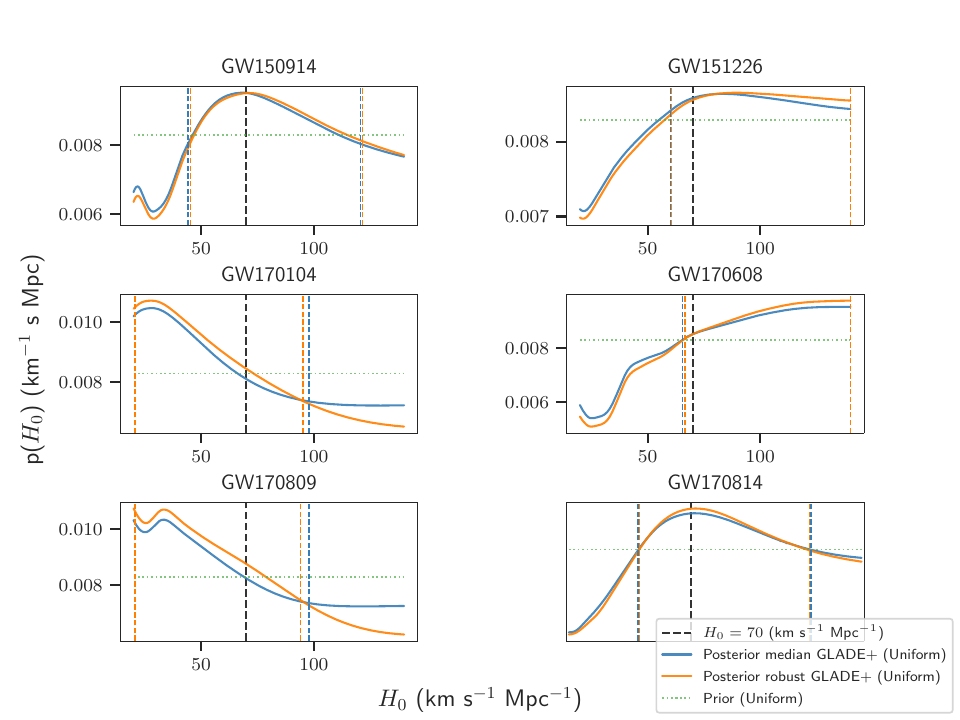}
    \caption{Results using the robust (blue) and median (orange) methods for individual events using the GLADE+ $B$-band with the GWTC-1 catalogue. The event GW170814 is analysed using the $g$-band of the DES-Y1 catalogue.}
    \label{fig:GWTC1+_all}
\end{figure*}

Figure \ref{fig:GW170814_DES} shows the posterior on $H_{0}$ for the event GW170814. The $87$ deg$^{2}$ sky localisation area of GW170814 was entirely contained within the DES footprint \citep{Doctor_2019}. The DES-Y1 catalogue consists of $\sim$137 million objects over $\sim$1800 deg$^{2}$ in the DES $grizY$ filters. The $10\sigma$ limiting magnitudes for galaxies are $g = 23.4$, $r = 23.2$, $i=22.5$, $z=21.8$ and $Y=20.1$ \citep{DESY1}. The catalogue includes photometric redshift estimates. The gravitational wave event GW170814 originated from within the area mapped by the DES-Y1 survey. In the analysis of GW170814 with \texttt{gwcosmo}, we use the $g$-band data from DES, as it is the band in which the survey is most complete, with the $95\%$ completeness magnitude limit in the $g$- band quoted at $23.72$ mag in a sample of high quality objects \citep{DESrelease}. 
Figure \ref{fig:GWTC1+_all} shows posteriors on $H_{0}$ for each individual dark siren in the GWTC-1 catalogue.

\subsection{GWTC-3}

The GWTC-3 catalogue consists of 90 events, of which 47 are used in this analysis. Of these 47 events, 42 are BBHs, 2 are BNSs (the bright siren GW170817
and GW190425), 2 are NSBHs (GW200105 and GW200115) and one is the asymmetric mass
binary GW190814 \citep{2023PhRvX..13d1039A,gw170817,GW190425,NSBH,GW190814}. In the original analysis presented in \cite{Abbott_2023}, the $K$-band of the GLADE+ catalogue was found to be more appropriate than the $B$-band for analysis; it is less affected by galactic dust, and the behaviour of its luminosity function can therefore be better approximated.  While this does not affect tests of galaxy catalogue completeness, it does affect the luminosity weighting of galaxies in the sample \citep{Abbott_2023}.

The completeness of the GLADE+ catalogue decreases more rapidly past $d_{L}\sim 100$Mpc in the $K$-band than in the $B$-band \citep{10.1093/mnras/sty1703}. This is because the $K$-band data comes from the 2MASS survey, which is not as deep as the newer surveys providing the $B$- and $W1$- band data. Applying the robust method to the GLADE+ catalogue with a pixel size of $N_{\rm side}=32$, the mean $K$-band apparent magnitude threshold is $m_{\rm thr} = 13.49$. By comparison, the median method gives  $m_{\rm thr} = 12.91$. While the robust method allows us to use more galaxies, the apparent magnitude threshold is still comparatively bright, reducing the impact of the method on the recovered $H_{0}$ compared to when using the $B$-band.

\begin{figure}
    \centering    \includegraphics[width=.5\textwidth]{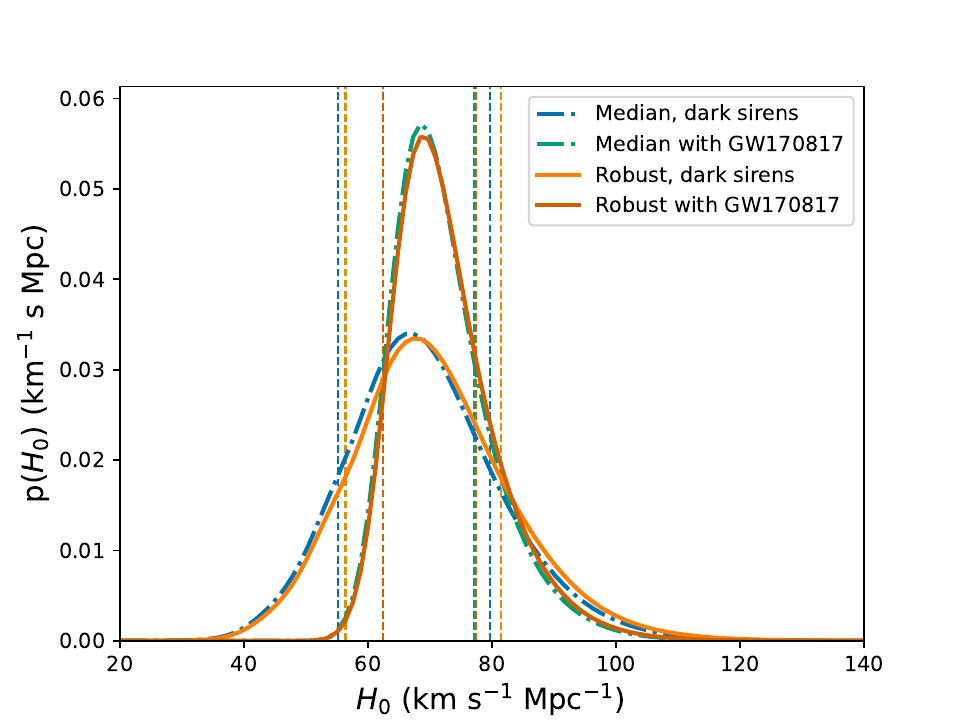}
    \caption{Final posterior on $H_{0}$ using the GLADE+ $K$-band on the GWTC-3 dataset. Vertical dashed lines show the $1\sigma$ intervals. The solid lines show results showing the robust method, while dash-dotted lines show results using $m_{\rm med}$ as the apparent magnitude threshold.}
    \label{fig:GWTC3+}
\end{figure}

\begin{figure*}
    \centering
    \includegraphics[width=\textwidth]{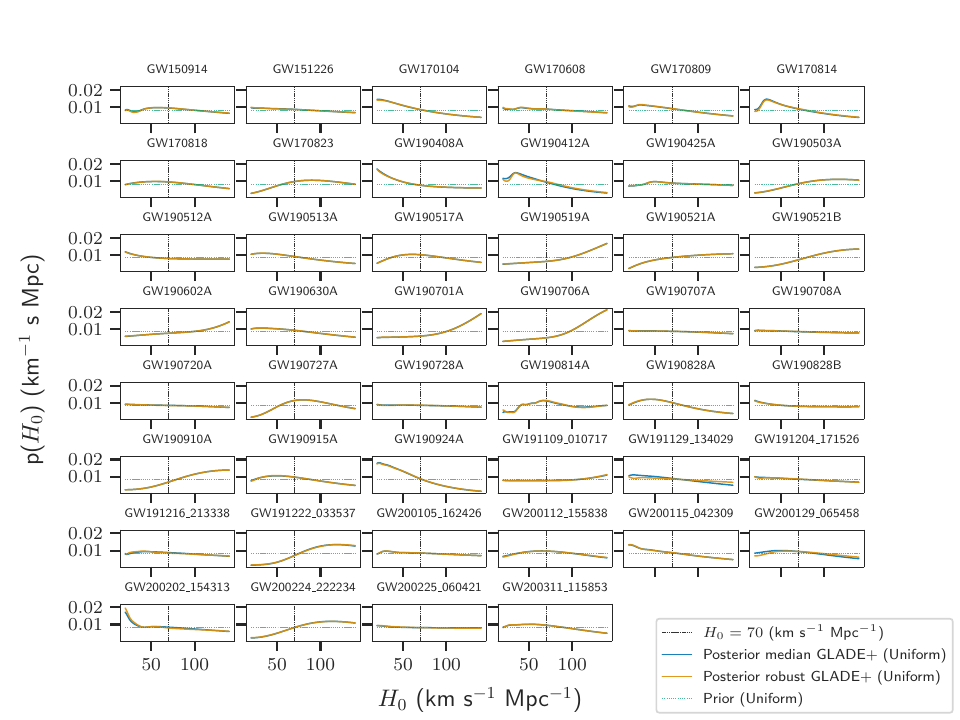}
    \caption{Inferred $H_{0}$ using the robust (orange) and median (blue) methods for estimating galaxy catalogue completeness, for individual events in the GWTC-3 catalogue. The $K$-band of the GLADE+ catalogue is used for analysis.}
    \label{fig:GWTC3+_all}
\end{figure*}

Results from the GWTC-3 analysis are shown in figure \ref{fig:GWTC3+}. As anticipated, there is no improvement in the recovered posterior on $H_{0}$ from this analysis. Figure \ref{fig:GWTC3+_all} shows results for individual event posteriors for both the robust and median methods. The recovered posteriors are similar for all events, due to the lack of coverage in the $K$-band at the luminosity distances at which potential host galaxies would be located. 

\begin{figure}
    \centering
    \includegraphics[width=.5\textwidth]{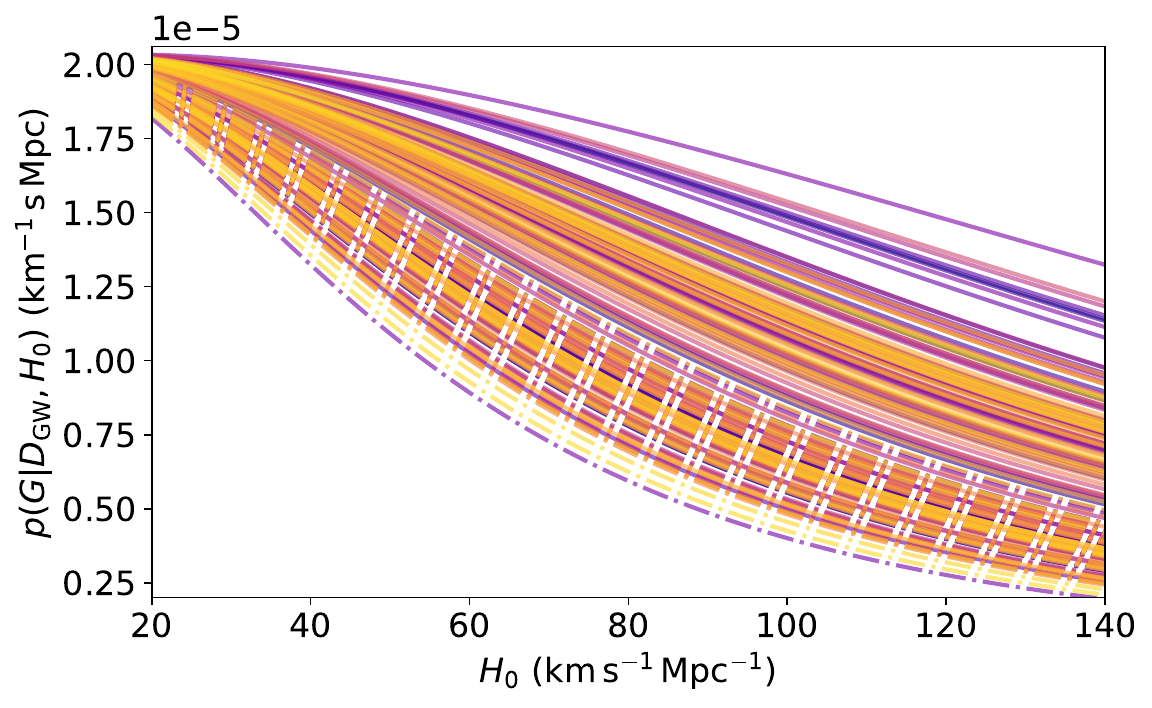}
    \caption{Probability of the host galaxy of the GW source being in the catalogue, $p(G|D_{\rm GW},H_{\rm 0})$, for each pixel of the skymap corresponding to the event GW190814A. The dash-dotted lines show $p(G|D_{\rm GW},H_{\rm 0})$ for the median method while the solid lines show $p(G|D_{\rm GW},H_{\rm 0})$ for the robust method. Even where events are uninformative, it is clear that using the robust method increases the contribution from $p(G|D_{\rm GW},H_{\rm 0})$.}
    \label{fig:GW190814A_pG}
\end{figure}

Figure \ref{fig:GW190814A_pG} illustrates the impact of the robust method on the probability of the host galaxy being within the galaxy catalogue for the event GW190814A using the GLADE+ $K$-band. Having a fainter apparent magnitude threshold increases the total contribution of the term $p(G|D_{\rm GW},H_{\rm 0})$ to the final posterior. In this case, however, where there is little galaxy catalogue support for the event, the contribution is still small compared to the $p(\overline{G}|D_{\rm GW},H_{\rm 0})$ term.

The final results for both GWTC-1 and GWTC-3 are summarised in table \ref{table:Results}.

\begin{table}
\vspace{0.5cm}
\begin{center}
\begin{tabular}{|l|l|p{24mm}|p{24mm}|}
\hline
Catalogue & Method & $H_{0}$ (km s$^{-1}$ Mpc$^{-1}$) & $H_{0}$ \footnotesize{(km s$^{-1}$ Mpc$^{-1}$)}\\
\hline
\multicolumn{2}{|c|}{} & Dark Sirens & with GW170817\\
\Xhline{2pt}

\multicolumn{2}{|c|}{GWTC-1, $B$-band}&\multicolumn{2}{|c|}{}\\
\hline
GLADE & median & $68.8^{+46.9}_{-21.7}$ & $69.6^{+19.2}_{-8.3}$ \\
GLADE & robust & $68.7^{+43.1}_{-24.6}$ & $69.5^{+18.1}_{-8.5}$ \\

\hline
GLADE+ & median & $65.7^{+41.8}_{-22.7}$ & $69.3^{+17.2}_{-8.3}$ \\
GLADE+ & robust & $67.9^{+35.1}_{-23.8}$ & $69.4^{+16.1}_{-7.8}$ \\

\Xhline{2pt}
\multicolumn{2}{|c|}{GWTC-3, $K$-band}&\multicolumn{2}{|c|}{}\\
\hline
GLADE+ & median & $66.8^{+12.9}_{-11.6}$& $68.6^{+8.4}_{-6.2}$\\
GLADE+ & robust & $67.7^{+13.7}_{-11.4}$& $68.9^{+8.5}_{-6.5}$ \\

\hline
\end{tabular}
\end{center}
\caption[GWTC-1 results]{Final results for constraints on $H_{0}$ in km s$^{-1}$ Mpc$^{-1}$ from the GWTC-1 and GWTC-3 datasets, using different methods and galaxy catalogues. The GWTC-1 dataset was analysed using the $B$-band of both the GLADE and GLADE+ catalogues, while the GWTC-3 dataset was analysed using the $K$-band of the GLADE+ catalogue. Confidence intervals are quoted at the $1\sigma$ level.}
\label{table:Results}
\end{table}

\section{Conclusions}
\label{sec:conc}

In this work we presented new results on the constraints on $H_{0}$ from dark sirens when we apply a robust test of completeness to the galaxy catalogue method.

There was no improvement to the posterior on $H_{0}$ with the robust method for the GWTC-3 analysis using the $K$-band of GLADE+. This is because the galaxy catalogue provides little or no coverage for any of the events in that band, whether the median or robust method is used. The final result is similar to the "empty catalogue" posteriors for each GWTC-3 event.

The final posterior on $H_{0}$ showed minor improvement when applying the robust method to the GWTC-1 analysis using the $B$-band of the GLADE+ catalogue. When only dark sirens were considered, the $1\sigma$ posterior was 8.6\% narrower when using the robust method than when using the median apparent magnitude as a threshold. While the GLADE+ $B$-band is less reliable for tests of completeness due to the behaviour of its luminosity function\footnote{See the discussion in \citep{Abbott_2023}}, the result demonstrates the need for a careful treatment of the apparent magnitude threshold of future, deeper galaxy catalogues in order to obtain the best constraints on $H_{0}$ from dark standard sirens.

The robust method applied here is more computationally expensive than simply taking the median apparent magnitude as a threshold --- with the complexity scaling as $\mathcal{N}_{\rm gal}^{2}$, where $\mathcal{N}_{\rm gal}$ is the number of galaxies in the sample. However, future instances of the pipeline will compute $m_{\rm thr}$ for the entire catalogue prior to analysis, circumventing the need to re-apply the method for each event. This will lead to improved performance and would eliminate the need for sub-sampling of galaxies. The threshold value used for determining $m_{\rm thr}$ from $T_{C}$ can also be refined in future work. Moreover, an uncertainty in the estimate of $m_{\rm thr}$ for each pixel can also be derived from the measurement uncertainties on apparent magnitudes and redshifts. These threshold uncertainties could then, in principle, also be incorporated into the \texttt{gwcosmo} pipeline. In this paper, the method was implemented into \texttt{gwcosmo} 1.0.0, but it will also be possible to incorporate it into version 2.0.0 and above.

While the robust method does not require that we know the exact form of the luminosity function, it does still make the assumption that the luminosity function is universal for the galaxy catalogue band considered. This represents a caveat when applying the robust method to the $B$-band of the GLADE and GLADE+ galaxy catalogues, and further investigation of the validity of this assumption for other bands and other catalogues will be carried out in future work.

Ongoing work ahead of the fifth LVK observing run, O5, is exploring the quantitative effects of having a deeper apparent magnitude limit in galaxy catalogues used for gravitational wave cosmology. With deeper surveys, we predict that excessively conservative estimates for $m_{\rm thr}$ will have a greater impact, making the implementation of robust completeness methods increasingly important in future work. Mock data challenges with deeper EM galaxy catalogues ahead of O5 will allow us to quantify the effect of using robust for future analyses.

Our future work will focus on applying the robust method to mock data in order to fully characterise potential biases and explore the effect of a more rigorously and robustly defined $m_{\rm thr}$ on the inference of $H_{0}$ when analysing GW data with deeper galaxy catalogues. We will also extend analysis to other colour bands, including the $B$ band, seeking to exploit the property of the robust method that it does not require the adoption of a specific parametric form for the galaxy luminosity function. Moreover, our future work we will also extend the analysis presented here to the case of galaxy surveys described by both a faint and bright apparent magnitude limit, applying the robust completeness test first developed in \cite{2007MNRAS.376.1757J}.

\section*{Acknowledgements}
The authors are grateful to En-Tzu Lin, Rachel Gray, Gavin Lamb, Surojit Saha, Surhud More, Maciej Bilicki, Gergely Dalya, Carl-Johan Haster and Suvodip Mukherjee for their helpful feedback and suggestions. They are also grateful to the referees for their helpful suggestions to improve this work.

This material is based upon work supported by NSF’s LIGO Laboratory which is a major facility fully funded by the National Science Foundation. L. D. was supported by the Science and Technology Facilities Council (Ref. ST/R504750/1), Nicholas and Lee Begovich and the Dan Black Family, and the National Science Foundation (Award 2308985). M. H. is supported by the Science and Technology Facilities Council (Ref. ST/L000946/1).  We acknowledge the use of the following python packages in this work: \texttt{gwcosmo} \citep{Gray2020}, Matplotlib \citep{hunter2007matplotlib}, healpy \citep{2005ApJ...622..759G,Zonca2019}.
This work has made use of CosmoHub, developed by PIC (maintained by IFAE and CIEMAT) in collaboration with ICE-CSIC. It received funding from the Spanish government (grant EQC2021-007479-P funded by MCIN/AEI/10.13039/501100011033), the EU NextGeneration/PRTR (PRTR-C17.I1), and the Generalitat de Catalunya.

\section*{Data Availability}

The GWTC-3 dataset is available from \cite{ligo_scientific_collaboration_and_virgo_2023_8177023}. The GWTC-1 dataset is available from \href{https://www.gw-openscience.org/GWTC-1}{https://www.gw-openscience.org/GWTC-1}. The GLADE and GLADE+ catalogues are available from the GLADE website \href{http://glade.elte.hu/}{http://glade.elte.hu/}. The DES-Y1 catalogue is available from \href{https://des.ncsa.illinois.edu/releases/y1a1}{https://des.ncsa.illinois.edu/releases/y1a1}.

\appendix
\section{Supplementary Material}
\label{sec:appendix}

\begin{figure}
    \centering
    \includegraphics[width=.5\textwidth]{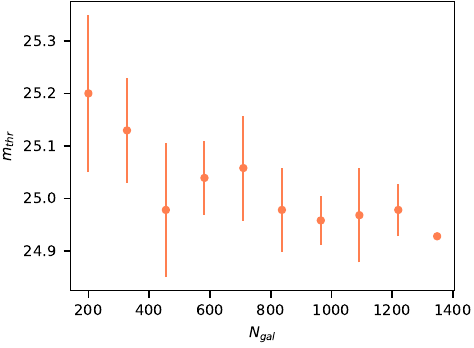}
    \caption{Calculated $m_{\rm thr}$ for one pixel of the MICE catalogue in the $r$-band at $N_{\rm side}$ = 32, against $N_{\rm gal}$, the number of galaxies in the sub-sample.}
    \label{fig:Ngal_ms}
\end{figure}

\begin{figure}
    \centering
    \includegraphics[width=.49\textwidth]{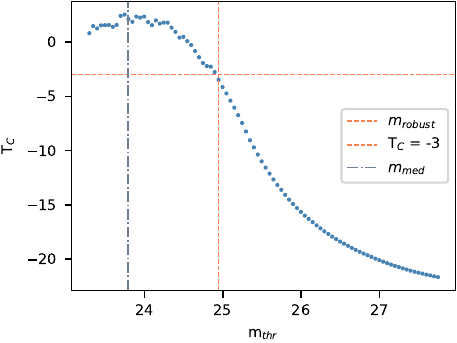}
    \caption{$T_{C}$ against trial apparent magnitude thresholds, for a sub-sample of 10000 MICECAT galaxies in the $r$-band. Following the methodology outlined in R01, the robust magnitude threshold $m_{\rm robust}$ is taken at $T_{C} = -3$, where $T_{C}$ starts to go strongly and systematically negative.}
    \label{fig:TCs}
\end{figure}

This appendix presents supplementary material illustrating the impact of some of the choices made in applying the robust test of incompleteness.
Figure~\ref{fig:Ngal_ms} shows $m_{\rm thr}$ for different $\mathcal{N}_{\rm gal}$ when sub-sampling galaxies contained in a healpy pixel with $N_{\rm side} = 32$ of the MICECAT catalogue. Figure~\ref{fig:TCs} shows, for a sample of 10000 MICECAT galaxies, the statistic $T_{C}$ against trial apparent magnitude thresholds.

\begin{figure}
    \centering
    \includegraphics[width=.49\textwidth]{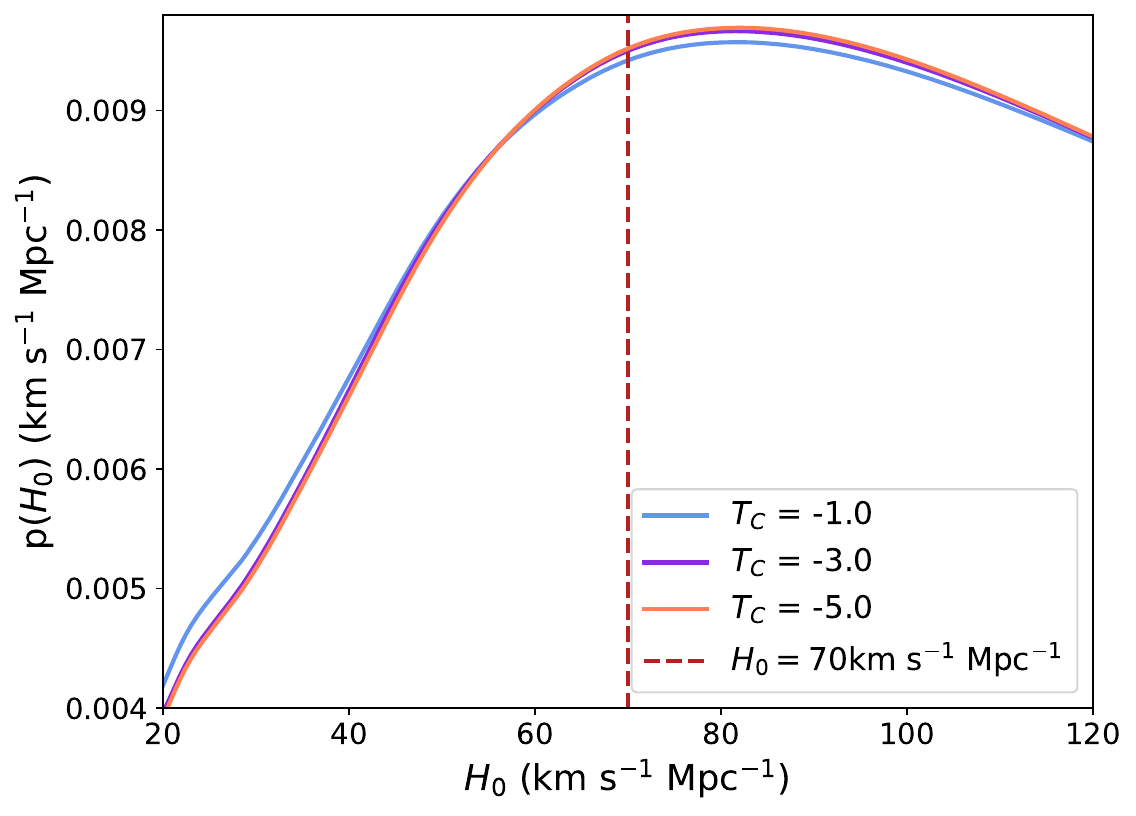}
    \caption{The posterior on GW150914 with different values for $T_{C}$, using the non-pixelated ("patch of sky" method) \texttt{gwcosmo} statistical pipeline with the GLADE+ $B$-band. There is a greater impact to choosing too high a value for $T_{C}$ than too low a value, since $T_{C}$ decreases steeply once it reaches apparent magnitudes at which the catalogue is incomplete.}
    \label{fig:TCs_GW150914}
\end{figure}

\begin{figure}
    \centering
    \includegraphics[width=.49\textwidth]{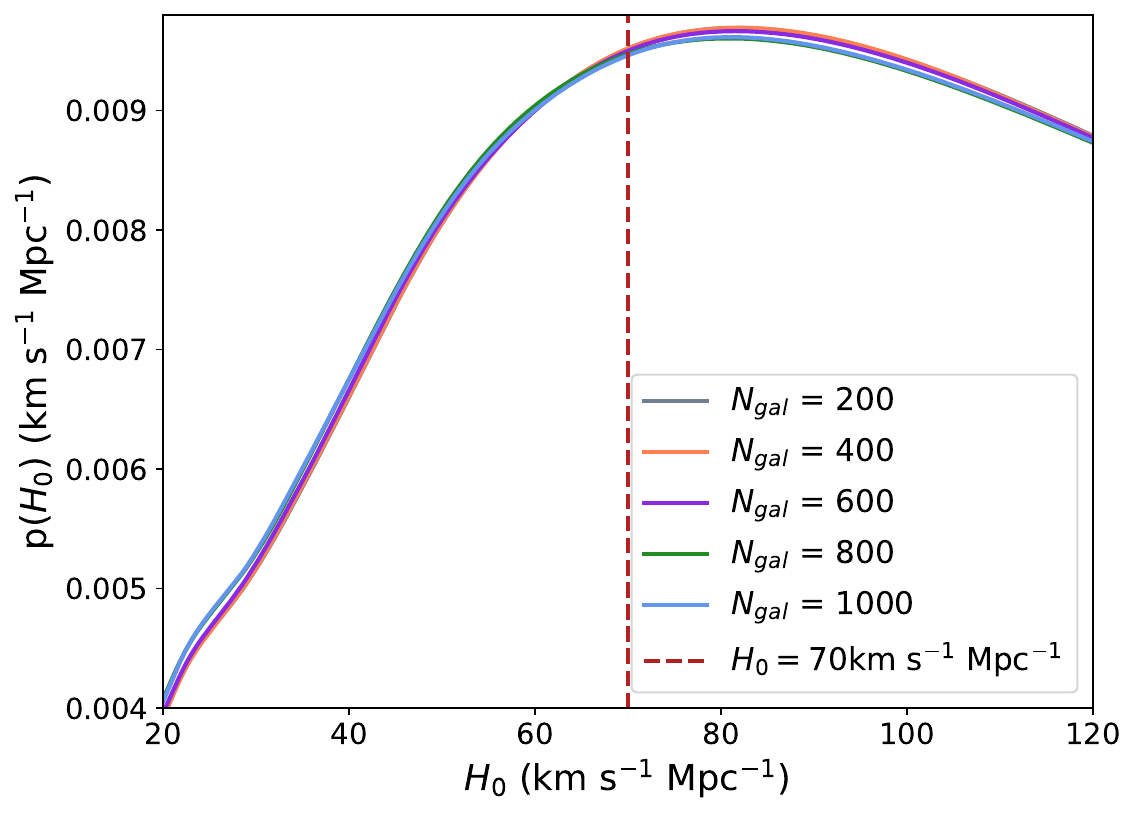}
    \caption{The posterior on GW150914 with different values for $N_{\rm gal}$, using the non-pixelated ("patch of sky" method) \texttt{gwcosmo} statistical pipeline with the GLADE+ $B$-band. }
    \label{fig:Ngal_GW150914}
\end{figure}

Figure~\ref{fig:TCs_GW150914} shows the posterior on $H_{0}$ for the event GW150914 using the non-pixelated \texttt{gwcosmo} pipeline and different $T_{C}$ thresholds with the GLADE+ B-band. Similarly, figure~\ref{fig:Ngal_GW150914} shows the $p(H_{0})$ posterior on GW150914 using different $\mathcal{N}_{\rm gal}$. We note that the full patch of sky for this analysis contains 708928 galaxies in the GLADE+ $B-$band.

\bibliographystyle{mnras}
\bibliography{bibli}

\bsp	
\label{lastpage}
\end{document}